\documentclass[]{spie}  
\usepackage{epsf,graphicx}
\usepackage{epsfig,color}
\usepackage{subfigure}

\title{The C-Band All-Sky Survey: Instrument design, status, and first-look data} 

\author{
Oliver G. King\supit{a*}, Charles Copley\supit{b,g}, Rod Davies\supit{c}, Richard Davis\supit{c}, Clive Dickinson\supit{c}, Yaser A. Hafez\supit{d}, Christian Holler\supit{e}, Jaya John John\supit{b}, Justin L. Jonas\supit{f}, Michael E. Jones\supit{b}, J. Patrick Leahy\supit{c}, Stephen J.C. Muchovej\supit{a}, Timothy J. Pearson\supit{a}, Anthony C.S. Readhead\supit{a}, Matthew A. Stevenson\supit{a}, and Angela C. Taylor\supit{b}
\skiplinehalf
\supit{a}California Institute of Technology, 1200 E. California Blvd., Pasadena 91125, USA \\
\supit{b}Univ. of Oxford, Denys Wilkinson Bldg. Keble Road, Oxford OX1 3RH, UK \\
\supit{c}Jodrell Bank Ctr. for Astrophysics, Univ. of Manchester M13 9PL, UK \\
\supit{d}KACST, P.O. Box 6086, Riyadh 11442, Saudi Arabia\\
\supit{e}Hochschule Esslingen, Kanalstra{\ss}e 33, Esslingen 73728, Germany\\
\supit{f}Rhodes Univ., Dept. Physics and Electronics, Grahamstown 6140, South Africa\\
\supit{g}Hartebeesthoek Radio Astronomy Observatory, P.O.Box 443, Krugersdorp 1740, South Africa
}

\authorinfo{* E-mail: ogk@astro.caltech.edu, Telephone: 1 626 395 4197}

\begin{document} 
  \maketitle 

\begin{abstract}
The C-Band All-Sky Survey (C-BASS) aims to produce sensitive, all-sky maps of diffuse Galactic emission at 5 GHz in total intensity and linear polarization. These maps will be used (with other surveys) to separate the several astrophysical components contributing to microwave emission, and in particular will allow an accurate map of synchrotron emission to be produced for the subtraction of foregrounds from measurements of the polarized Cosmic Microwave Background. We describe the design of the analog instrument, the optics of our 6.1~m dish at the Owens Valley Radio Observatory, the status of observations, and first-look data.
\end{abstract}



\section{INTRODUCTION} \label{sec:intro}

The fluctuations in the Cosmic Microwave Background (CMB) contain a vast amount of information on the history and composition of our universe. In particular, the angular power spectra of the fluctuations contain rich detail about our early universe, significantly constraining the cosmological parameters in our current models of the universe. The latest cosmological data firmly support an inflationary $\Lambda$CDM cosmology for our Universe, with a minimal 6 parameter model (the ``concordance cosmology'')\cite{Komatsu:2009p1317}. The next challenge in cosmology is to probe the physics of the inflationary period by looking for the signature of primordial gravitational waves in the polarized CMB.

CMB polarization was generated at last scattering by perturbations in the primordial fluid, caused by scalar and tensor perturbations\cite{Baumann:2008p1184}. The tensor perturbations result from the ``stretching'' of space by gravitational wave fluctuations, while scalar perturbations result from the same density fluctuations in the primordial fluid that result in the CMB temperature anisotropy. The ratio of the tensor perturbation amplitude to the scalar perturbation amplitude, $r$, is a key tracer of the physics of the inflationary epoch.

A local quadrupole anisotropy in the radiation field at the time of decoupling would cause the CMB in that region to become very slightly linearly polarized (of order $\sim10^{-6}$), through Thomson scattering by electrons\cite{Hu:2003p1726,Rees:1968p30}. The linear polarization of the CMB is described by the Stokes parameters $Q$ and $U$. However, $Q$ and $U$ are orientation dependent, and hence produce a rotationally-variant polarization power spectrum. The CMB ($Q$,$U$) polarization can be decomposed\cite{Kamionkowski:1997p1294,Zaldarriaga:1997p247} into two rotationally invariant quantities, called E and B. The E- and B-modes are often described as the curl-free and divergence-free modes of the polarization vector respectively. The CMB B-mode is a direct tracer of the tensor perturbations caused by gravitational waves in the inflationary period of the universe. It has consequently been dubbed the ``smoking gun'' of inflation.

However, the Galactic foreground emission can have any mixture of E- and B-modes. Measurements of the E- and B-modes of the \emph{WMAP} polarization spectra show that the CMB B-mode is obscured by the B-mode of polarized Galactic synchrotron emission at all angular scales when averaged over the whole sky\cite{Page:2007p205}. There are, however, small ($\sim 10^{3}$ square degree) low-emission regions of the sky where the CMB polarization is expected to dominate.

The early small-area, high-latitude, experiments which measured the EE power spectrum of the CMB\cite{Readhead:2004p817,Barkats:2005p5,Leitch:2005p773,Montroy:2006p776,Sievers:2007p780} focused on low foreground emission areas of the sky where foreground subtraction was not necessary. However, a robust detection of the much-fainter B-mode power spectrum is likely to require data from large areas of the sky and hence some form of foreground removal, particularly if the tensor-to-scalar ratio $r$ is small ($\leq 0.1$).

\begin{figure}
 \centering
\subfigure[]
{
 \includegraphics[width=0.45\textwidth]{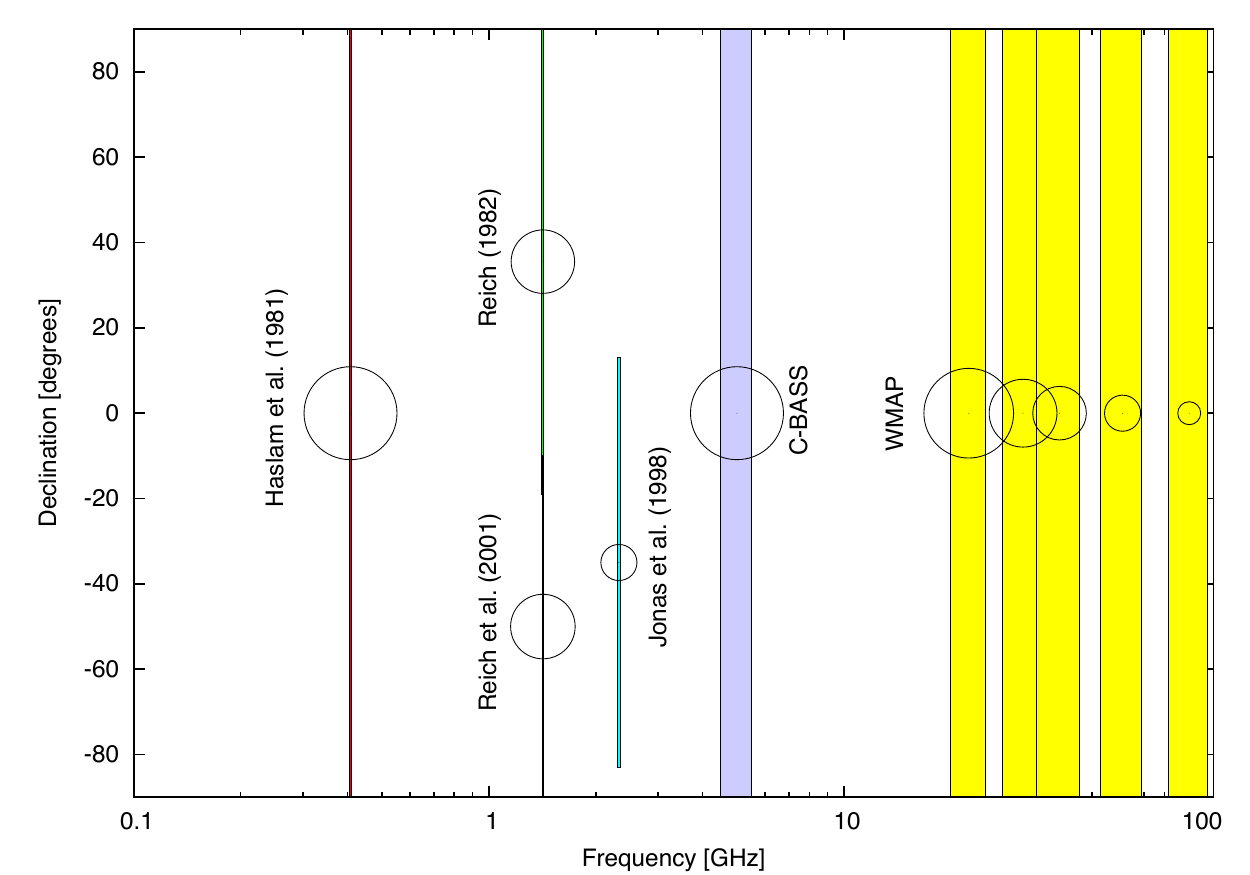}
 \label{fig:I_sky_surveys}
}
\subfigure[]
{
 \includegraphics[width=0.45\textwidth]{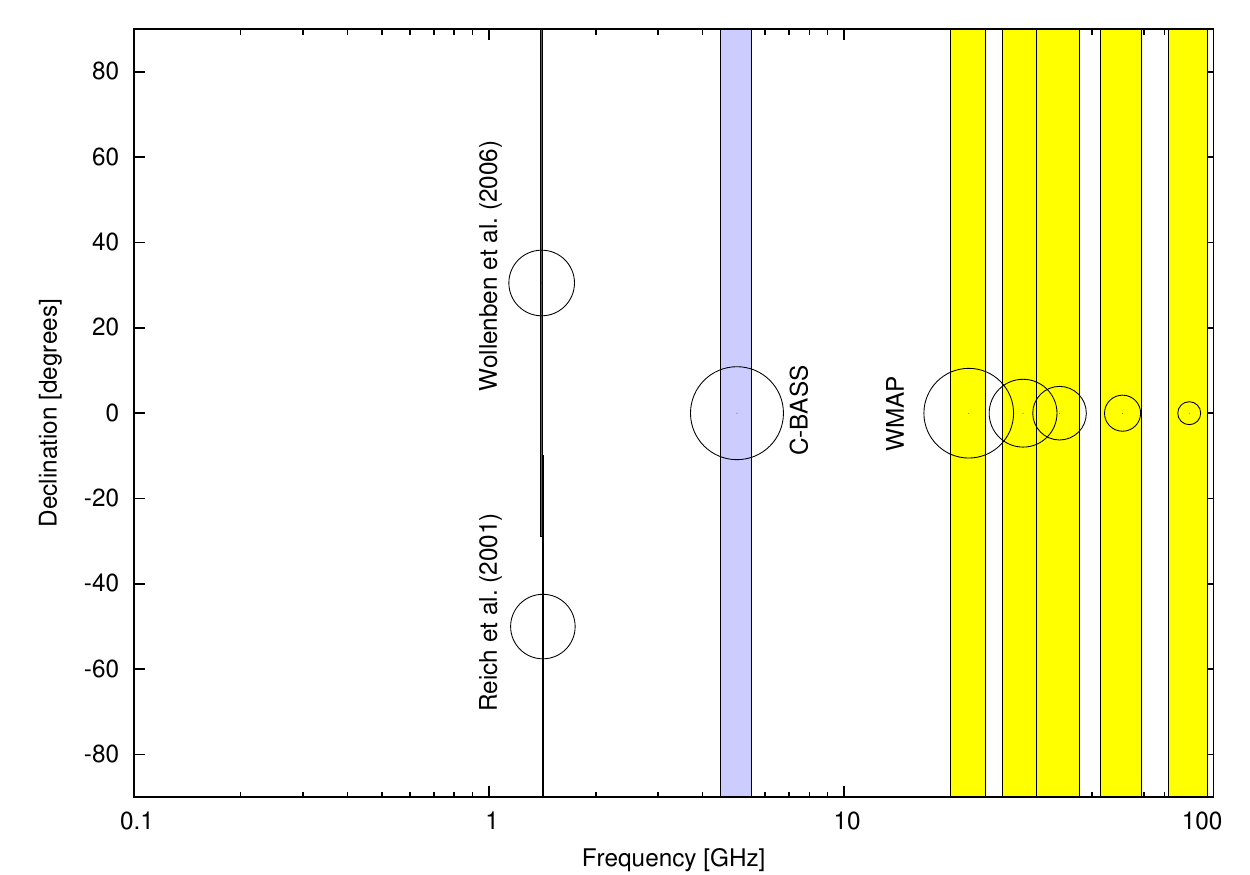}
 \label{fig:P_sky_surveys}
}
\caption{Existing large-area radio, microwave, and mm surveys between 0.1 and 100~GHz, with C-BASS shown for comparison. (a) Total intensity\cite{Haslam:1981,Reich:1982,Reich:2001p19,Jonas:1998,Gold:2009p339}; (b) Polarization\cite{Reich:2001p19,Wolleben:2005p151,Gold:2009p339}. The vertical axis shows sky coverage in declination; coverage is complete in right ascension. The horizontal axis is frequency, with the width of a survey box indicating the frequency coverage, analogous to sensitivity. The diameter of the circle is proportional to the angular resolution of the survey; for reference the C-BASS angular resolution is shown as 0.85$^{\circ}$.}
\label{fig:sky_surveys}
\end{figure}

The dominant source of polarized emission at CMB frequencies accessible from the ground is Galactic synchrotron emission. This emission is from high-energy charged particles spiraling around Galactic magnetic field lines, and is strongly linearly polarized. The polarized Galactic synchrotron emission is currently poorly characterized; the current status of moderate-resolution, large sky area surveys is shown in Figure~\ref{fig:sky_surveys}. The \emph{WMAP} 23~GHz polarization map\cite{Gold:2009p339} has a low signal-to-noise ratio over much of the sky, while the 1.4~GHz polarization surveys\cite{Reich:2001p19,Wolleben:2005p151} are expected to be badly affected by Faraday rotation at low Galactic latitudes.

C-BASS aims to provide sensitive, all-sky maps of both total intensity and linear polarization at 5~GHz. This frequency is high enough that we will not be unduly affected by Faraday rotation over most of the sky, and low enough that we can reach a good sensitivity in a reasonable amount of time. In addition to being useful for polarized foreground subtraction from CMB measurements, C-BASS will provide a wealth of data on the nature of the Galactic magnetic field, firm constraints on polarized ``anomalous'' emission, and additional data on the Galactic cosmic ray electron population. C-BASS will allow, for the first time, an accurate separation of synchrotron, free-free, and anomalous emission. In particular, it will constrain any flatter spectrum components that could be responsible for both the \emph{WMAP} haze and anomalous dust emission.

\section{RECEIVER DESIGN} 

\begin{figure}
 \centering
 \includegraphics[width=0.9\textwidth]{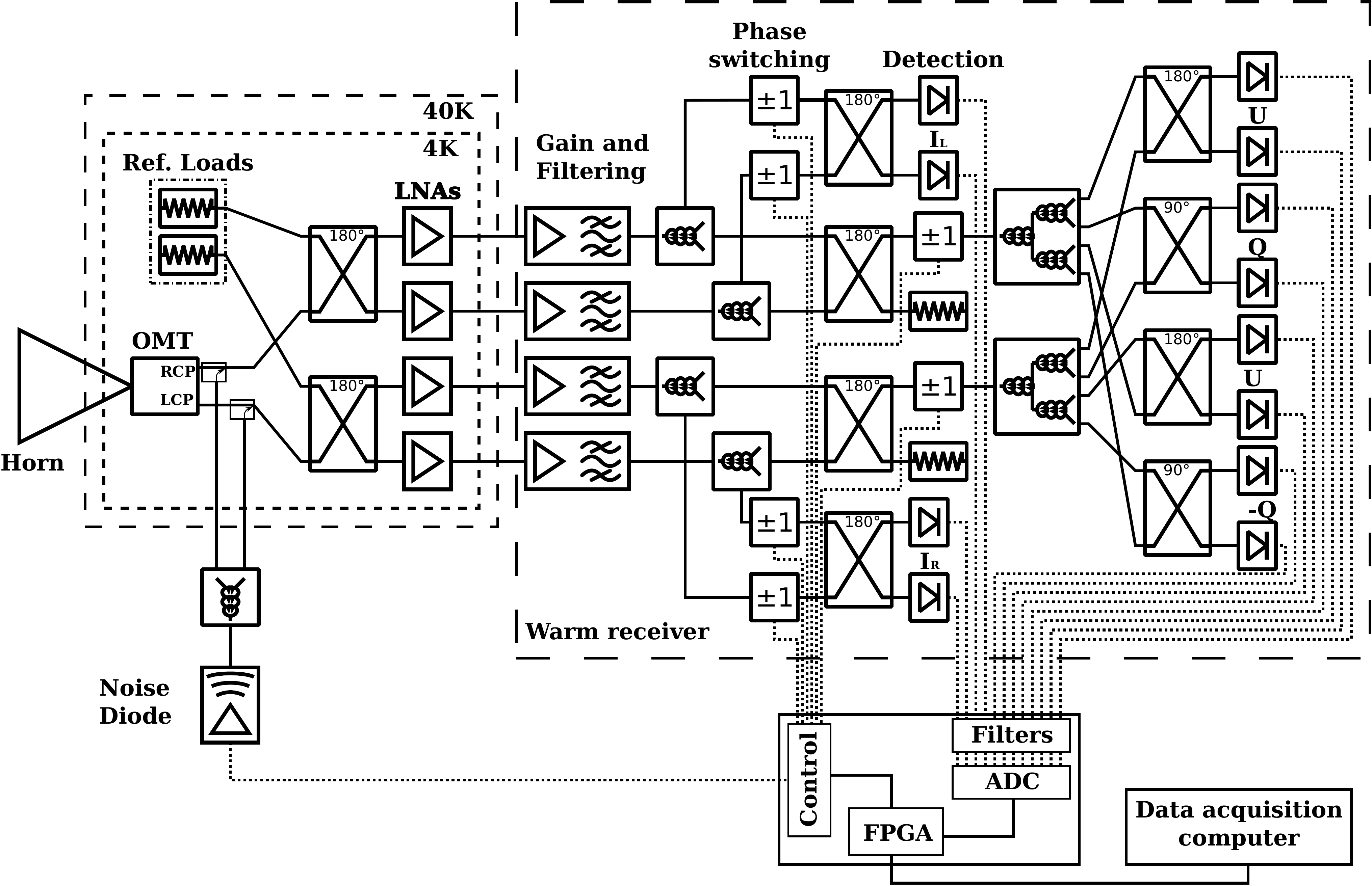}
\caption{Simplified schematic layout of the C-BASS receiver. The horn is followed by an OMT which produces orthogonal circular polarizations. A calibration noise signal is injected equally into the circular polarizations. These are then combined with reference load signals using $180^{\circ}$ hybrids, followed by LNAs. These signals exit the cryostat and are further amplified and filtered by band-defining filters. Further separation and combination of RF signals is performed before the RF signals are detected using detector diodes. The detector diode voltages are digitized and further processed by an FPGA.}
 \label{fig:receiver_layout}
\end{figure}

The C-BASS receiver is a novel hybrid of a pseudo-correlation radiometer and a pseudo-correlation polarimeter, covering the 4.5 to 5.5~GHz band in a single analog channel. A schematic diagram of the receiver is shown in Figure~\ref{fig:receiver_layout}. This receiver architecture was chosen to provide stable measurements of both total intensity (Stokes $I$) and linear polarization (Stokes $Q$ and $U$). The detector diode voltages are digitized and processed in a custom-designed digital backend.

Total intensity is measured by pseudo-correlating both orthogonal circular polarizations against independent reference loads. This style of radiometer has been used in both the \emph{WMAP}\cite{Jarosik:2003p204} and \emph{Planck}\cite{Davis:2009p1768} receivers. Like the \emph{Planck} receiver, we compare orthogonal circular polarization signals against electrical refence loads, rather than signals from a second horn \emph{\`{a} la} \emph{WMAP}. This allows us to preserve the optical purity of the receiver by having an on-axis optical system. The electrical reference loads are temperature controlled RF terminations. Suppression of receiver gain drifts, given unequal sky and reference load temperatures, can be improved through the use of an r-factor correction\cite{Mennella:2003p2064}.

We expect that the diffuse Galactic synchrotron emission will have a negligible circularly polarized component. In this case, each orthogonal circular polarization measured by the receiver will be a noise-independent measure of the sky signal total intensity. The pseudo-correlation radiometer provides a more stable measurement of total intensity, at the expense of increased receiver noise.

The receiver measures both Stokes $Q$ and $U$ simultaneously in a pseudo-correlation fashion. When $E_{l}$ and $E_{r}$, the left-circular and right-circular polarization voltages respectively, are pseudo-correlated using a $180^{\circ}$ hybrid we obtain Stokes $U$. When they are pseudo-correlated with a $90^{\circ}$ hybrid we obtain Stokes $Q$. In the C-BASS receiver $E_{l}$ and $E_{r}$ are pseudo-correlated in all possible combinations to provide two measures of each linear polarization parameter.

We now discuss some aspects of the receiver in greater detail.

\subsection{Optics}

The C-BASS receiver will be installed on telescopes at sites in both the Northern and Southern hemispheres. The telescopes and chosen sites are a 6.1~m diameter dish designed by JPL as a prototype for the DSN replacement network\cite{Imbriale:2005p1510}, located at the Owens Valley Radio Observatory (OVRO), near Big Pine in California; and a 7.6~m diameter former satellite communications antenna located at the MeerKAT support site, near Carnarvon in South Africa.

\begin{figure}
 \centering
 \includegraphics[width=0.7\textwidth]{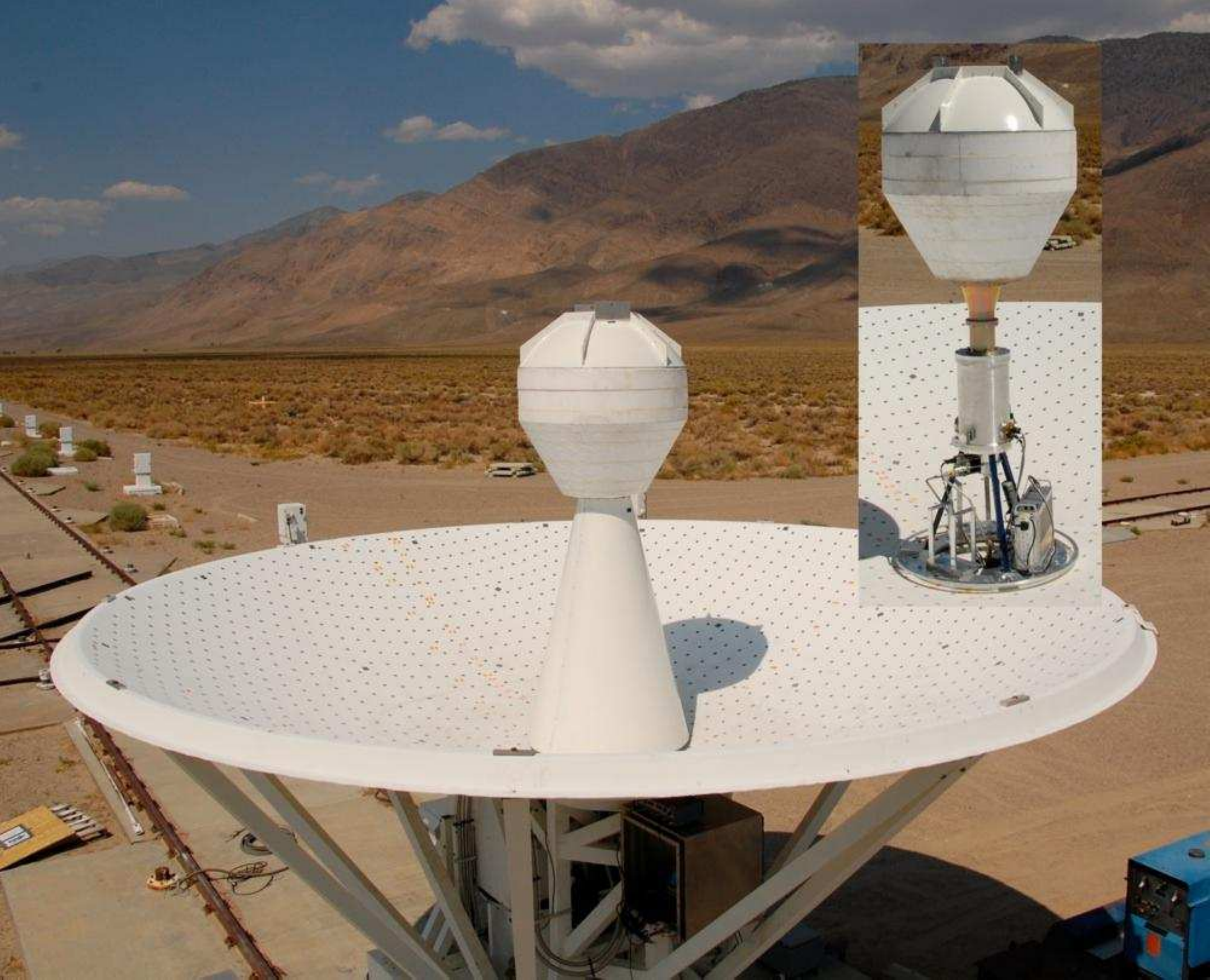}
\caption{Photograph of the C-BASS dish at OVRO, showing the foam cone secondary support. The inset shows the interior of the metal shroud, which houses the cryostat and warm receiver. The absorbing tunnels around the primary and secondary mirrors had not yet been installed at the time this photograph was taken.}
 \label{fig:dish_assembly}
\end{figure}

A significant feature of the optical systems of both telescopes is that the secondary mirrors are not supported by metal legs running up from the dish surface \cite{Holler:2010}. Metal support legs scatter the incoming wave and introduce an unwanted spurious polarization to the signal and uncontrolled signal contamination by ground pick-up. The secondary mirrors are, instead, supported on a cone of radio-transparent foam fixed to the receiver shroud, as shown in Figure~\ref{fig:dish_assembly}. The foam was Plastazote LD45\footnote{http://zotefoams.com/pages/EN/datasheets/LD45.HTM}, a closed-cell, cross-linked, nitrogen-inflated polyethylene foam. Stress modelling indicates that the foam is rigid enough that only 0.92~mm of deflection is seen when it is tipped at an angle of $45^{\circ}$.

\begin{figure}
 \centering
\subfigure[]
{
 \includegraphics[width=0.45\textwidth]{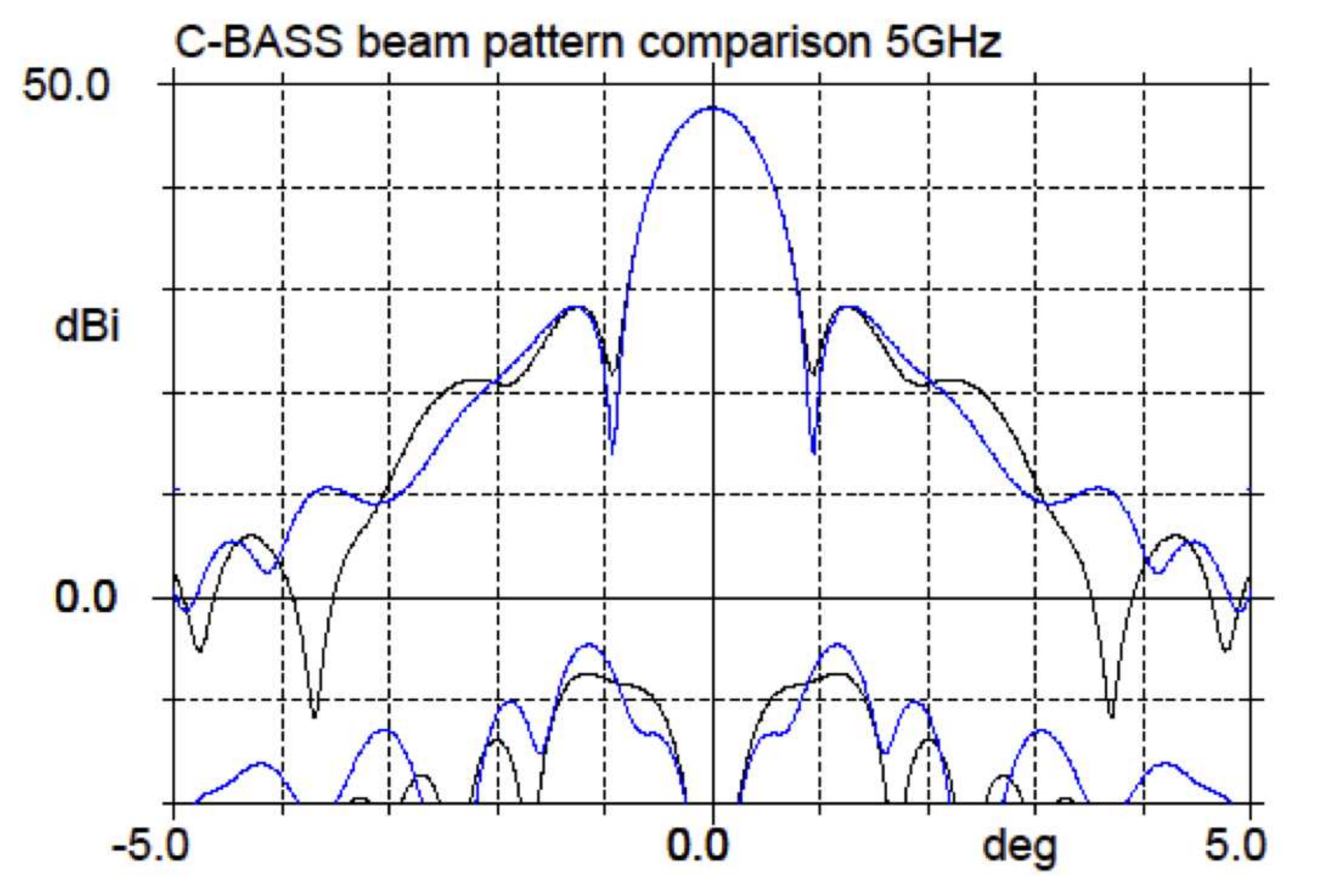}
\label{fig:beam_shape_comparison}
 }
\subfigure[]
{
 \includegraphics[width=0.45\textwidth]{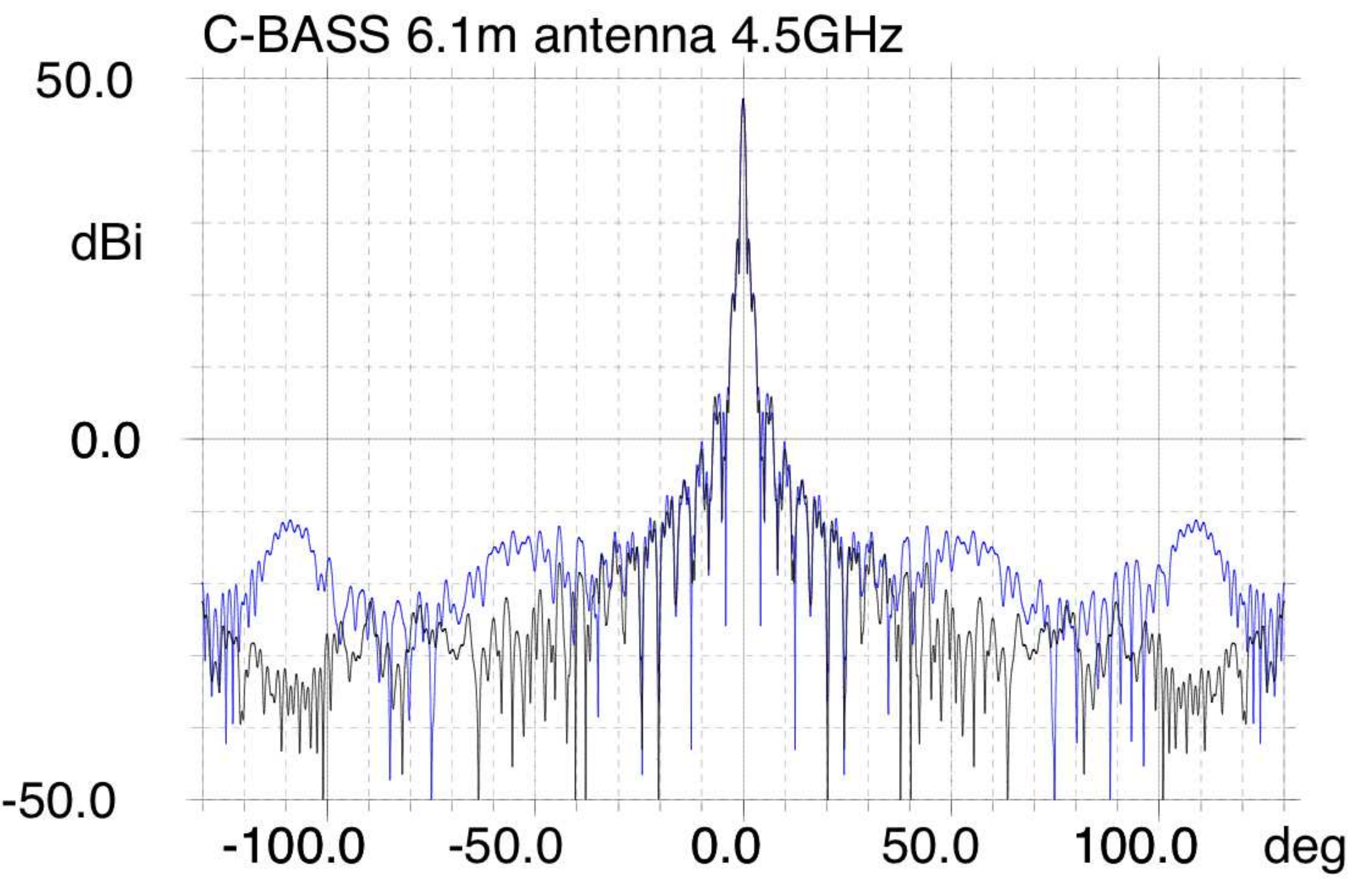}
\label{fig:tunnel_effect}
}
\caption{(a) The simulated E-plane beam shapes and cross-polarizations of the 6.1~m diameter JPL dish (black trace) and the 7.6~m diameter South African dish (blue trace)\cite{Holler:2010}. Note the identical main-beam shape and excellent cross-polarization level of $<-50$~dB. The simulations were performed in GRASP. (b) Simulated E-plane beam shapes for the JPL dish without (blue) and with (black) absorbing tunnels. Note the reduction of the far-out sidelobes by the absorbing tunnel.}
\label{fig:optics}
\end{figure}

Both telescopes have shaped, i.e. non-parabolic, primary mirrors and were originally equipped with unsuitable secondary mirrors and horns. The new optical systems use the same profiled corrugated horn with new custom secondary mirrors. The secondary mirrors are shaped such that both telescopes have the same beam shape, as shown in Figure~\ref{fig:beam_shape_comparison}. In the JPL dish the secondary has a Gregorian form, while the South African secondary has a Cassegrain form. The beams have a HPBW of about $0.85^{\circ}$.

A ray-trace analysis revealed spillover past the secondary mirror. In order to reduce the resulting sidelobe, and a second far-out sidelobe, absorbing tunnels will be placed around the primary and secondary of the JPL dish. This is not necessary for the South African dish, as we under-illuminate the larger primary. The simulated beam pattern of the JPL dish, with absorbing tunnels around both the primary and secondary, is shown in Figure~\ref{fig:tunnel_effect}. We see a significant reduction in the far-out sidelobes at the lower end of our band.


\subsection{Polarimeter}

The C-BASS receiver, as shown in Figure~\ref{fig:receiver_layout}, is divided into three sections: the cold receiver, the warm receiver, and the digital readout system. The cold receiver consists of the LNAs and pre-LNA components -- kept cold (at $\sim$4~K) to keep the noise penalty of any loss low -- housed in a cryostat. The warm receiver is housed in a separate box beneath the cryostat and consists of all the components from after the LNAs to the detector diodes. The readout system filters and digitizes the detector diode outputs.

The C-BASS cryostat is ``dry'', using a Sumitomo Heavy Industries SRDK-408D2 cold head to provide cooling. The cold head provides two stages of cooling: $\sim45$~W at 40~K and $\sim1.0$~W at 4.2~K. The cryostat consists of an outer cylindrical body which forms part of the mechanical structure of the telescope, an inner 40~K heat shield to reduce thermal loading on the second stage, and a 4~K second stage.

\begin{figure}
 \centering
\subfigure[]
{
 \includegraphics[width=0.7\textwidth]{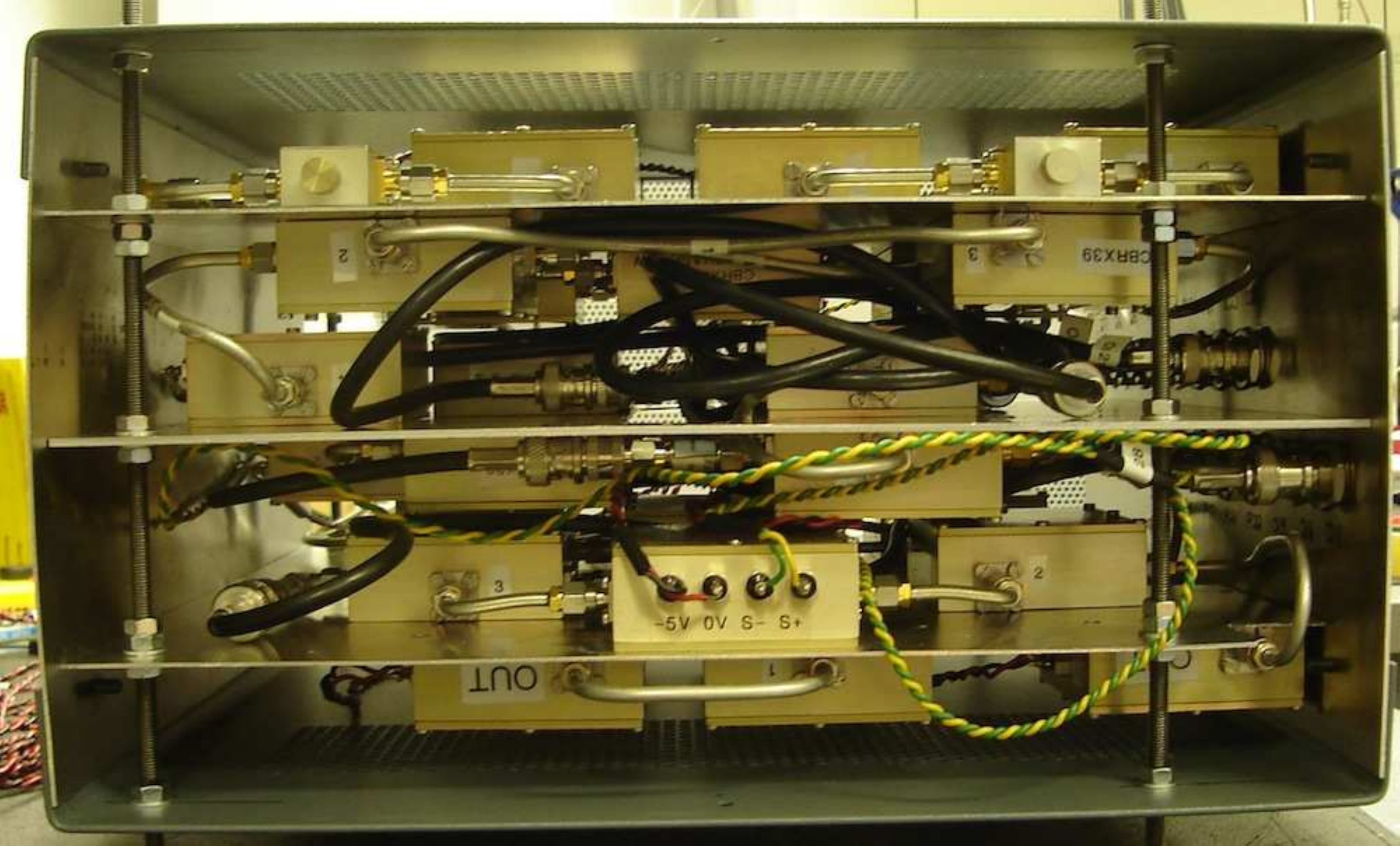}
 \label{fig:warm_receiver_side_view}
}
\subfigure[]
{
 \includegraphics[width=0.25\textwidth]{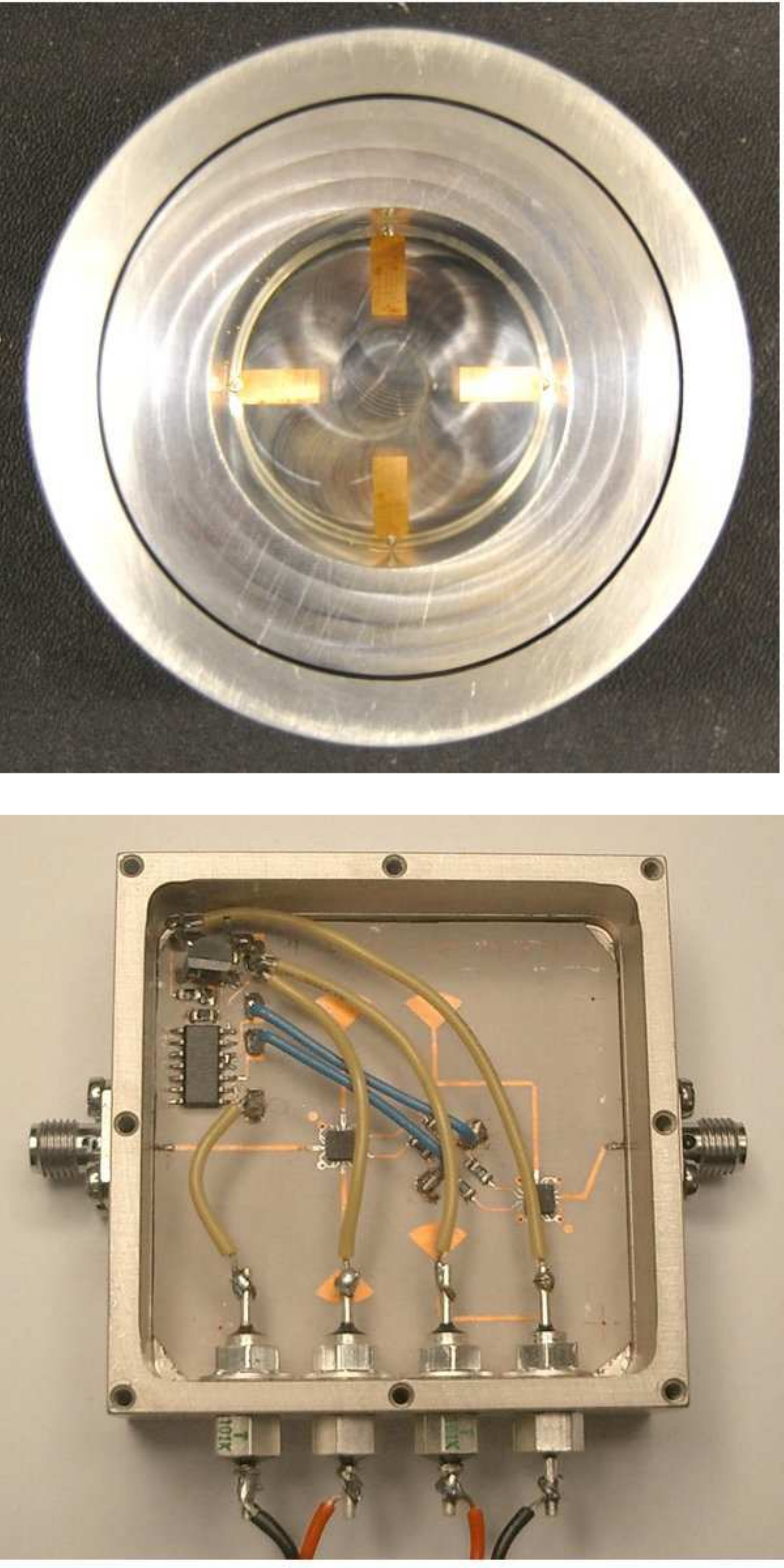}
 \label{fig:novel_parts}
}
\caption{(a) Side-view of the warm receiver box. The four RF signals from the LNAs enter through the left face. The detector diode outputs and phase switch control signals are piped through BNC connections on the right face. The RF components are arranged in three layers to take up as little space as possible in the small area between the cryostat and dish surface. (b) Some of the novel components developed for the C-BASS receiver: the orthomode transducer (top) and $180^{\circ}$ phase switch (bottom).}
\label{fig:warm_polarimeter_parts}
\end{figure}

The warm receiver, consisting of the components between the LNAs and the detector diodes, was implemented as a single analog channel covering the 4.5 to 5.5~GHz band. An image of the warm receiver is shown in Figure~\ref{fig:warm_receiver_side_view}. This box is housed in the space between the cryostat and the dish surface (see Figure~\ref{fig:dish_assembly}). The broad fractional bandwidth, limited budget, and stringent performance requirements required us to develop a number of novel components. Among these were a compact, broad-band, and low cross-polarization orthomode transducer\cite{Grimes:2006p1688}, and novel broad-band $180^{\circ}$ phase switches, shown in Figure~\ref{fig:novel_parts}.

\subsection{Digital Readout}

The digital backend performs post-detection processing of the analog channels in preparation for archiving by the control computer.  Two cards are used: an analog filterboard, and an ADC/FPGA-based, digital card.  The ADC board was designed for the LiCAS particle physics experiment\cite{Reichold:2006}, but after some modifications to the filters and a re-programming of the FPGA, it met the C-BASS requirements.

As well as post-detection processing, the backend is also responsible for control of the phase switching signals and the noise diode.  The processing itself is comprised of low-pass filtering, digitization, demodulation, integration, and preparation for transfer to the control computer.  The last three steps are performed on the Xilinx Spartan-3 FPGA, and data are transferred to the computer via USB.

\section{FIRST-LOOK DATA}

The C-BASS project is currently in a hybrid observing mode. Commissioning and active development of the data reduction pipeline are ongoing. We have begun test survey observations to supply data to the data reduction pipeline and map making routines.

\begin{figure}
 \centering
\subfigure[]
{
 \includegraphics[width=0.45\textwidth]{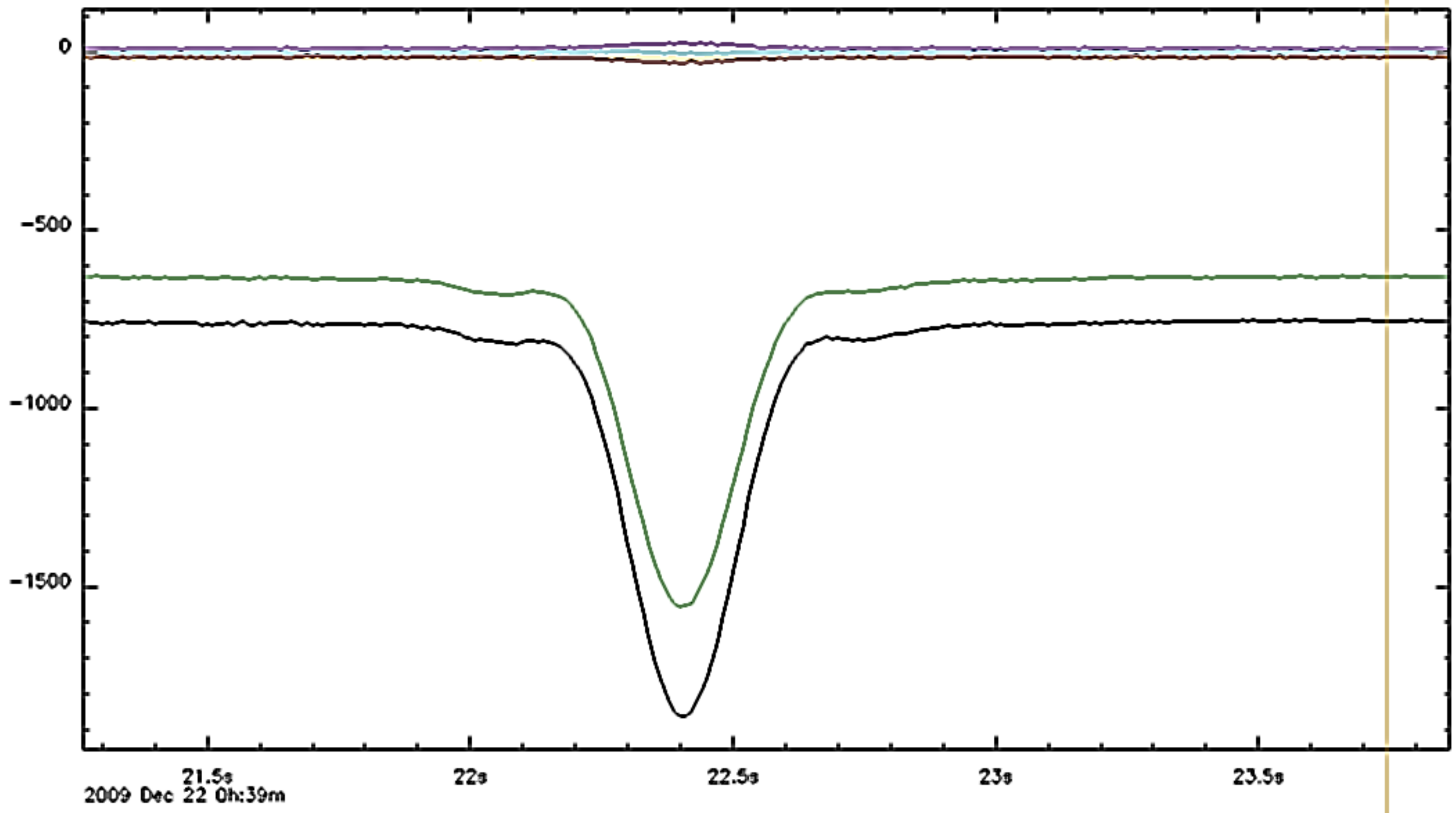}
 \label{fig:theMoon}
}
\subfigure[]
{
 \includegraphics[width=0.45\textwidth]{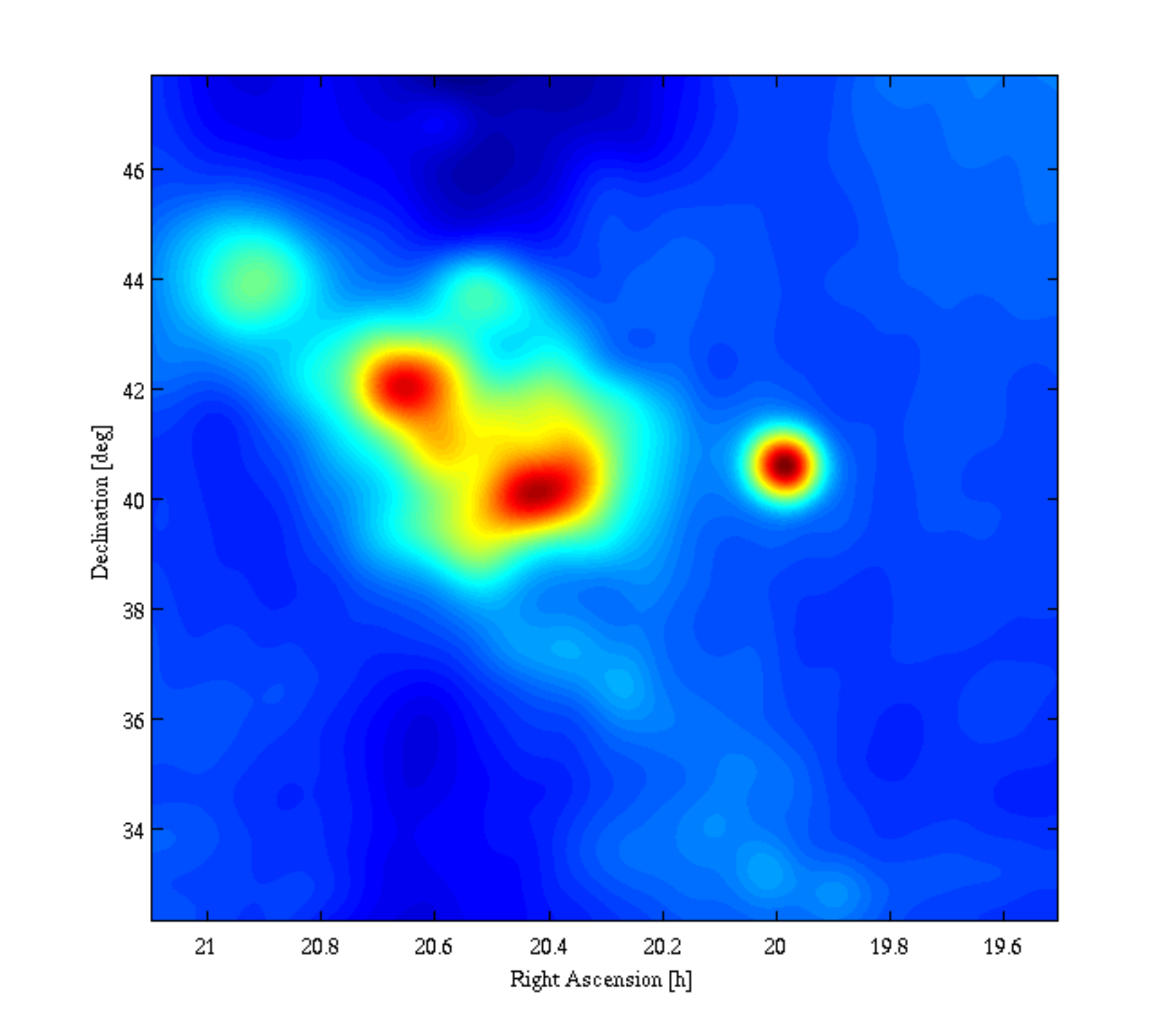}
 \label{fig:cygamap}
}
\caption{(a) First light on the C-BASS receiver in December 2009: raw detector diode outputs from a scan through the moon. The traces with a large variation are total intensity; the slightly-varying traces are linear polarization, uncorrected for instrumental leakage. (b) Map of the Cygnus region from May 2010, in uncalibrated total intensity. The compact source on the right is the radio galaxy Cygnus-A, which is dominated by synchrotron emission. The diffuse emission to the left of Cygnus-A is the Cygnus-X region of the Galactic plane, which is dominated by free-free emission.}
\end{figure}

First light on the C-BASS receiver was achieved in December 2009. The raw detector diode outputs from a scan through the moon are shown in Figure~\ref{fig:theMoon}. The traces which show a large fluctuation are total intensity. The remaining traces are linear polarization, uncorrected for instrumental leakage.

We present, in Figure~\ref{fig:cygamap}, a map of the Cygnus region in uncalibrated total intensity. The compact source on the right is the radio galaxy Cygnus-A, which is dominated by synchrotron emission. The diffuse emission to the left of Cygnus-A is the Cygnus-X region of the Galactic plane, which is dominated by free-free emission\cite{Davies:1957}. The image was made by raster scanning the region with scans in azimuth. Linear baselines were removed from each azimuth scan, RFI-contaminated data was removed, and the data was binned on a sparse grid and then convolved with the C-BASS beam. The data were taken on three successive nights, with 1.6~hours of observing contributing to the final map.

\section{TIMELINE AND SECOND RECEIVER}

We plan to build a second receiver for C-BASS. It will cover the same band as the first receiver, but will use digital hardware to perform the necessary correlations. It will produce spectral measurements of the 4.5 to 5.5~GHz band for RFI mitigation purposes. The spectra will be collapsed into a single bin to match the data produced by the analog instrument.

\begin{figure}
 \centering
\includegraphics[width=\textwidth]{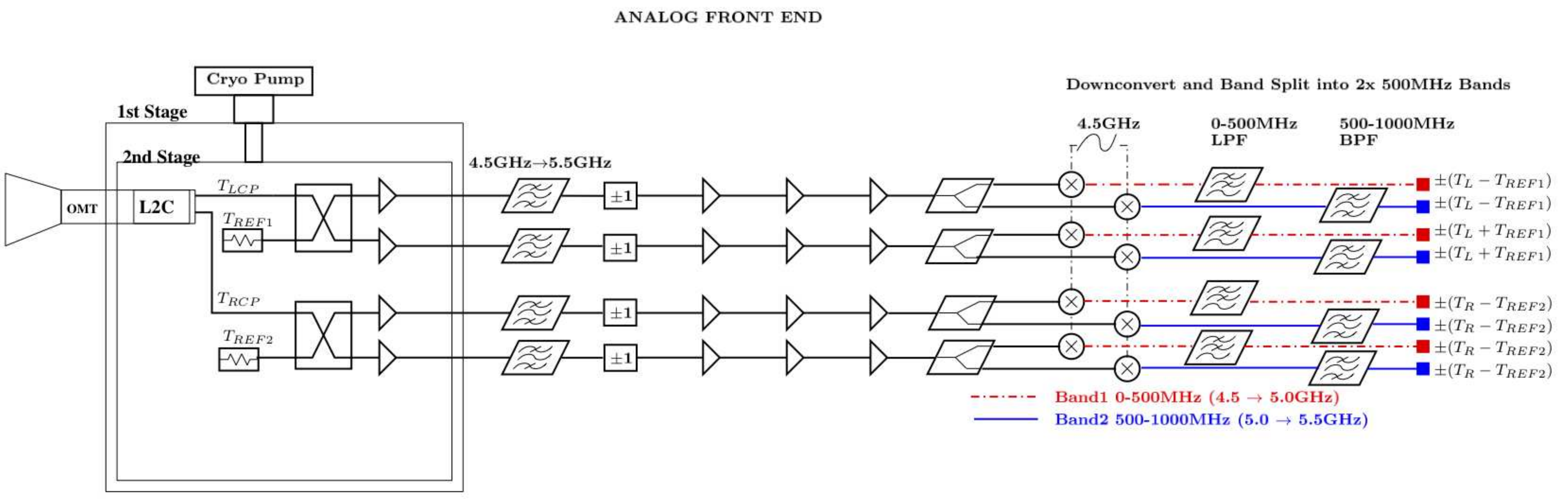}
\caption{Schematic diagram of the analog section of the second C-BASS receiver. Orthogonal circular polarizations from the horn are combined with reference load signals, amplified, phase switched, and filtered. The RF signals are mixed down to the DC to 500~MHz and 500~MHz to 1~GHz bands for digitization and processing by ROACH boards.}
\label{fig:2nd_receiver}
\end{figure}

A schematic diagram of the second C-BASS receiver is shown in Figure~\ref{fig:2nd_receiver}. Orthogonal circular polarizations from the horn are combined with reference load signals, amplified, phase switched, and filtered. The 4.5 to 5.5~GHz RF signals are mixed down to the DC to 0.5~GHz and 0.5 to 1~GHz bands for digitization and processing. The digital signal capture and processing will be performed on CASPER\footnote{http://casper.berkeley.edu/wiki/ROACH} hardware, specifically iADC analog to digital convertors and the Xilinx Virtex 5 based ROACH board. This design environment is able to take advantage of the robust and well supported CASPER software toolflow, facilitating both rapid design and deployment as well as remote reconfigurability.

We expect to move gradually from the present hybrid commissioning/observing mode to a dedicated observing mode over the next quarter. The first published data from the Northern survey are not expected until 2011, with the Southern survey data to follow in 2012.

\section{CONCLUSIONS}

The analog 4.5 to 5.5~GHz C-BASS receiver has been built and is installed at the Owens Valley Radio Observatory in California. A number of beam maps have been performed to verify the optical performance of the receiver. It is currently in a hybrid commissioning/observing mode, with the data reduction pipeline in active development. We present here a map in uncalibrated total intensity of the Cygnus region. We plan to build a second receiver with a digital spectral backend covering the same band for installation at the Southern telescope in South Africa. We expect to release the Northern survey data in 2011, with the Southern survey data to follow in 2012.

\acknowledgments     

The C-BASS project is a collaboration between Caltech/JPL in the US, Oxford and Manchester Universities in the UK, and Rhodes University and the Hartebeesthoek Radio Astronomy Observatory in South Africa.  It is funded by the NSF (AST-0607857) and the participating institutions. We would like to thank the Xilinx University Programme for their donation of FPGAs to this project. http://www.astro.caltech.edu/cbass/



\begin{thebibliography}{10}

\bibitem{Komatsu:2009p1317}
Komatsu, E., Dunkley, J., Nolta, M.~R., Bennett, C.~L., Gold, B., Hinshaw, G.,
  Jarosik, N., Larson, D., Limon, M., Page, L., Spergel, D.~N., Halpern, M.,
  Hill, R.~S., Kogut, A., Meyer, S.~S., Tucker, G.~S., Weiland, J.~L., Wollack,
  E., and Wright, E.~L., ``Five-year Wilkinson microwave anisotropy probe
  observations: Cosmological interpretation,'' {\em ApJS}~{\bf 180},  330 (2009).

\bibitem{Baumann:2008p1184}
Baumann, D., Jackson, M.~G., Adshead, P., Amblard, A., Ashoorioon, A., Bartolo,
  N., Bean, R., Beltran, M., de~Bernardis, F., Bird, S., Chen, X., Chung, D.
  J.~H., Colombo, L., Cooray, A., Creminelli, P., Dodelson, S., Dunkley, J.,
  Dvorkin, C., Easther, R., Finelli, F., Flauger, R., Hertzberg, M.,
  Jones-Smith, K., Kachru, S., Kadota, K., Khoury, J., Kinney, W.~H., Komatsu,
  E., Krauss, L.~M., Lesgourgues, J., Liddle, A., Liguori, M., Lim, E., Linde,
  A., Matarrese, S., Mathur, H., McAllister, L., Melchiorri, A., Nicolis, A.,
  Pagano, L., Peiris, H.~V., Peloso, M., Pogosian, L., Pierpaoli, E., Riotto,
  A., Seljak, U., Senatore, L., Shandera, S., Silverstein, E., Smith, T.,
  Vaudrevange, P., Verde, L., Wandelt, B., Wands, D., Watson, S., Wyman, M.,
  Yadav, A., Valkenburg, W., and Zaldarriaga, M., ``CMBPol mission concept
  study: Probing inflation with CMB polarization,'' {\em AIP Conf. Proc.}~{\bf 1141}, 10--120
  (2009).

\bibitem{Hu:2003p1726}
Hu, W., Hedman, M.~M., and Zaldarriaga, M., ``Benchmark parameters for CMB
  polarization experiments,'' {\em Phys. Rev. D}~{\bf 67},  43004 (2003).

\bibitem{Rees:1968p30}
Rees, M.~J., ``Polarization and spectrum of the primeval radiation in an
  anisotropic universe,'' {\em AJ}~{\bf 153},  L1 (1968).

\bibitem{Kamionkowski:1997p1294}
Kamionkowski, M., Kosowsky, A., and Stebbins, A., ``Statistics of cosmic
  microwave background polarization,'' {\em Phys. Rev. D (Particles)}~{\bf
  55},  7368 (1997).

\bibitem{Zaldarriaga:1997p247}
Zaldarriaga, M. and Seljak, U., ``All-sky analysis of polarization in the
  microwave background,'' {\em Phys. Rev. D (Particles)}~{\bf 55},  1830
  (1997).

\bibitem{Page:2007p205}
Page, L., Hinshaw, G., Komatsu, E., Nolta, M.~R., Spergel, D.~N., Bennett,
  C.~L., Barnes, C., Bean, R., Dor{\'e}, O., Dunkley, J., Halpern, M., Hill,
  R.~S., Jarosik, N., Kogut, A., Limon, M., Meyer, S.~S., Odegard, N., Peiris,
  H.~V., Tucker, G.~S., Verde, L., Weiland, J.~L., Wollack, E., and Wright,
  E.~L., ``Three-year Wilkinson microwave anisotropy probe (WMAP) observations:
  Polarization analysis,'' {\em ApJS}~{\bf 170},  335 (2007).

\bibitem{Readhead:2004p817}
Readhead, A. C.~S., Myers, S.~T., Pearson, T.~J., Sievers, J.~L., Mason, B.~S.,
  Contaldi, C.~R., Bond, J.~R., Bustos, R., Altamirano, P., Achermann, C.,
  Bronfman, L., Carlstrom, J.~E., Cartwright, J.~K., Casassus, S., Dickinson,
  C., Holzapfel, W.~L., Kovac, J.~M., Leitch, E.~M., May, J., Padin, S.,
  Pogosyan, D., Pospieszalski, M., Pryke, C., Reeves, R., Shepherd, M.~C., and
  Torres, S., ``Polarization observations with the cosmic background imager,''
  {\em Science}~{\bf 306},  836 (2004).

\bibitem{Barkats:2005p5}
Barkats, D., Bischoff, C., Farese, P., Fitzpatrick, L., Gaier, T., Gundersen,
  J.~O., Hedman, M.~M., Hyatt, L., McMahon, J.~J., Samtleben, D., Staggs,
  S.~T., Vanderlinde, K., and Winstein, B., ``First measurements of the
  polarization of the cosmic microwave background radiation at small angular
  scales from CAPMAP,'' {\em AJ}~{\bf 619},  L127 (2005).

\bibitem{Leitch:2005p773}
Leitch, E.~M., Kovac, J.~M., Halverson, N.~W., Carlstrom, J.~E., Pryke, C., and
  Smith, M. W.~E., ``Degree angular scale interferometer 3 year cosmic
  microwave background polarization results,'' {\em AJ}~{\bf 624},  10 (2005).

\bibitem{Montroy:2006p776}
Montroy, T.~E., Ade, P. A.~R., Bock, J.~J., Bond, J.~R., Borrill, J.,
  Boscaleri, A., Cabella, P., Contaldi, C.~R., Crill, B.~P., Bernardis, P.~D.,
  Gasperis, G.~D., Oliveira-Costa, A.~D., Troia, G.~D., Stefano, G.~D., Hivon,
  E., Jaffe, A.~H., Kisner, T.~S., Jones, W.~C., Lange, A.~E., Masi, S.,
  Mauskopf, P.~D., Mactavish, C.~J., Melchiorri, A., Natoli, P., Netterfield,
  C.~B., Pascale, E., Piacentini, F., Pogosyan, D., Polenta, G., Prunet, S.,
  Ricciardi, S., Romeo, G., Ruhl, J.~E., Santini, P., Tegmark, M., Veneziani,
  M., and Vittorio, N., ``A measurement of the CMB <EE> spectrum from the 2003
  flight of BOOMERanG,'' {\em AJ}~{\bf 647},  813 (2006).

\bibitem{Sievers:2007p780}
Sievers, J.~L., Achermann, C., Bond, J.~R., Bronfman, L., Bustos, R., Contaldi,
  C.~R., Dickinson, C., Ferreira, P.~G., Jones, M.~E., Lewis, A.~M., Mason,
  B.~S., May, J., Myers, S.~T., Oyarce, N., Padin, S., Pearson, T.~J.,
  Pospieszalski, M., Readhead, A. C.~S., Reeves, R., Taylor, A.~C., and Torres,
  S., ``Implications of the cosmic background imager polarization data,'' {\em
  AJ}~{\bf 660},  976 (2007).

\bibitem{Haslam:1981}
Haslam, C.~G., Klein, U., Salter, C.~J., Stoffel, H., Wilson, W.~E., Cleary, M.~N., Cooke, D.~J. and Thomasson, P., 
``A 408 MHz all-sky continuum survey. I - Observations at southern declinations and for the North Polar region,'' 
{\em A{\&}A}~{\bf 100}, 209 (1981).

\bibitem{Jonas:1998}
Jonas, J.~L., Baart, E. and Nicolson, G.,
``The Rhodes/HartRAO 2326-MHz radio continuum survey,''
{\em MNRAS}~{\bf 297}, 977 (1998).

\bibitem{Reich:1982}
Reich, W., ``A radio continuum survey of the northern sky at 1420 MHz. I.''
{\em A{\&}AS}~{\bf 48}, 219 (1982).

\bibitem{Gold:2009p339}
Gold, B., Bennett, C.~L., Hill, R.~S., Hinshaw, G., Odegard, N., Page, L.,
  Spergel, D.~N., Weiland, J.~L., Dunkley, J., Halpern, M., Jarosik, N., Kogut,
  A., Komatsu, E., Larson, D., Meyer, S.~S., Nolta, M.~R., Wollack, E., and
  Wright, E.~L., ``Five-year Wilkinson microwave anisotropy probe observations:
  Galactic foreground emission,'' {\em ApJS}~{\bf 180},  265--282 (2009).

\bibitem{Reich:2001p19}
Reich, P., Testori, J.~C., and Reich, W., ``A radio continuum survey of the
  southern sky at 1420 MHz. The atlas of contour maps,'' {\em A{\&}A}~{\bf
  376},  861 (2001).

\bibitem{Wolleben:2005p151}
Wolleben, M., Landecker, T.~L., Reich, W., and Wielebinski, R., ``An absolutely
  calibrated survey of polarized emission from the northern sky at 1.4 GHz,''
  {\em arXiv}~{\bf astro-ph/0510456} (2005).

\bibitem{Jarosik:2003p204}
Jarosik, N., Bennett, C.~L., Halpern, M., Hinshaw, G., Kogut, A., Limon, M.,
  Meyer, S.~S., Page, L., Pospieszalski, M., Spergel, D.~N., Tucker, G.~S.,
  Wilkinson, D.~T., Wollack, E., Wright, E.~L., and Zhang, Z., ``Design,
  implementation, and testing of the microwave anisotropy probe radiometers,''
  {\em ApJS}~{\bf 145},  413 (2003).

\bibitem{Davis:2009p1768}
Davis, R.~J., Wilkinson, A., Davies, R.~D., Winder, W.~F., Roddis, N.,
  Blackhurst, E.~J., Lawson, D., Lowe, S.~R., Baines, C., Butlin, M., Galtress,
  A., Shepherd, D., Aja, B., Artal, E., Bersanelli, M., Butler, R.~C.,
  Castelli, C., Cuttaia, F., D'Arcangelo, O., Gaier, T., Hoyland, R., Kettle,
  D., Leonardi, R., Mandolesi, N., Mennella, A., Meinhold, P., Pospieszalski,
  M., Stringhetti, L., Tomasi, M., Valenziano, L., and Zonca, A., ``Design,
  development and verification of the 30 and 44 GHz front-end modules for the
  Planck low frequency instrument,'' {\em Journal of Inst.}~{\bf 12},
   12002 (2009).

\bibitem{Mennella:2003p2064}
Mennella, A., Bersanelli, M., Seiffert, M., Kettle, D., Roddis, N., Wilkinson,
  A., and Meinhold, P., ``Offset balancing in pseudo-correlation radiometers
  for CMB measurements,'' {\em A{\&}A}~{\bf 410},  1089
  (2003).

\bibitem{Imbriale:2005p1510}
Imbriale, W. and Gama, E., ``Antennas for the array-based deep space network:
  current status and future designs,'' {\em Aerospace Conference, 2005 IEEE} ,
  1140 -- 1149 (2005).

\bibitem{Holler:2010}
Holler, C. {et al}, ``Circular symmetric antenna design with high polarization purity and low spillover,'' (in prep.).

\bibitem{Grimes:2006p1688}
Grimes, P., King, O., Yassin, G., and Jones, M., ``Compact broadband planar
  orthomode transducer,'' {\em Electronics Letters}~{\bf 43},  1146 -- 1147
  (2006).

\bibitem{Reichold:2006}
Reichold, A. for the LiCAS Collaboration, ``The LiCAS-RTSR - a survey system for the ILC,'' in
  [{\em Proceedings of the 10th biennial European Particle Accelerator
  Conference}{\nolinebreak\hspace{0.1em}]},
  http://cern.ch/AccelConf/e06/PAPERS/MOPCH195.PDF (2006).

\bibitem{Davies:1957}
Davies, R.~D., ``On the nature of the Cygnus-X radio source as derived from observations in the continuum and at the hydrogen-line frequency,'' {\em MNRAS}~{\bf 117}, 663 (1957).

\end{thebibliography}
\end{document}